\documentstyle[11pt]{article}

\input epsf

 \newcommand{\beq}{\begin{equation}}
\newcommand{\eeq}{\end{equation}} \newcommand{\beqn}{\begin{eqnarray}}
\newcommand{\eeqn}{\end{eqnarray}}

\begin{document}
\thispagestyle{empty}
\baselineskip=18pt
\rightline{KIAS-P99104}
\rightline{SNUTP-99-048}
\rightline{{\tt hep-th/9911186}}
\vskip 2cm
\centerline{\Large\bf  Quantum Spectrum of Instanton Solitons }
\centerline{\Large\bf in Five Dimensional  Noncommutative $U(N)$ Theories}

\vskip 0.2cm

\vskip 1.2cm
\centerline{\large\it
Kimyeong Lee \footnote{Electronic Mail: klee@kias.re.kr}
and Piljin Yi \footnote{Electronic Mail: piljin@kias.re.kr}
}
\vskip 5mm
\centerline{ \it School of Physics, Korea Institute for
Advanced Study}
\centerline{\it
207-43, Cheongryangri-Dong, Dongdaemun-Gu, Seoul 130-012, Korea}

\vskip 1.2cm
\begin{quote}
{\baselineskip 16pt We explore quantum states of instanton solitons in
five dimensional noncommutative Yang-Mills theories. We start with
maximally supersymmetric $U(N)$ theory compactified on a circle $S^1$,
and derive the low energy dynamics of instanton solitons, or calorons,
which is no longer singular. Quantizing the low energy dynamics, we
find $N$ physically distinct ground states with a unit Pontryagin
number and no electric charge. These states have a natural D-string
interpretation.  The conclusion remains unchanged as we decompactify
$S^1$, as long as we stay in the Coulomb phase by turning on adjoint
Higgs expectation values.  }
\end{quote}

\newpage

\section{Instantons and Monopoles in Noncommutative Yang-Mills}

In Yang-Mills theories in five dimensions, solitons carrying
Pontryagin charge, or real time instantons, can appear. Unlike their
four-dimensional cousins, magnetic monopoles, the low energy dynamics
of these solitons are singular due to well-known singularities in
instanton moduli spaces.  A singularity appears when any scale
parameter vanishes, i.e., when an instanton size becomes zero.  This
prevents a reliable computation of low energy spectra in sectors with
nonzero Pontryagin numbers.

 Recently, it was realized that a noncommutative version of such
Yang-Mills theories arises naturally from open string propagating on
D-branes~\cite{connes,doug,seiberg}. While we will not go into
details, it suffices to observe that the resulting theory can be
understood by allowing coordinates to be noncommuting variables. One
effect of this is to remove very small distance scales.  Therefore,
one expects to find that the instanton cannot be smaller than the 
size set by the noncommutativity.  Indeed, recent study of
noncommutative Yang-Mills theory indicates that small instanton
singularity~\cite{witten} of the instanton moduli space is resolved
once we deform the Yang-Mills theory in this
manner~\cite{nek,ber,nakajima,seiberg}. Naturally, this opens a way to
regularize dynamics of five-dimensional solitons.

Before delving into low energy dynamics, it is instructive to
understand how solitons themselves are deformed. Easiest way to
construct such deformed solitons on noncommutative $R^{1+4}$ would be
via the ADHM construction. In string theory context, one is
considering the D0 dynamics inside $N$ D4-branes. The Higgs vacua of
D0 parameterize the instanton moduli space of the $U(N)$ theory, which
can be found by solving a series of D-term conditions \cite{witten}.  
These D-term conditions are nothing but the ADHM equation~\cite{adhm}.

However, a little bit more intuitive picture is obtained by
considering T-dual of this when compactified on $ R^{1+3} \times S^1 $.
Let the radius of $S^1$ be $R$. One can take the decompactification
limit $R\rightarrow \infty$ at the end, if one wishes. The T-duality
maps D4-branes to D3-branes whose worldvolumes are parallel to
$R^{1+3}$ and transverse to the dual circle $\tilde S^1$ of radius
$\tilde{R}= \alpha'/R$. For $U(N)$ gauge theory on $ R^{1+3}\times
S^1$, one has $N$ such D3 branes. The instanton charge is given by the
D-string winding number along $\tilde S^1$, so a single instanton
corresponds to a singly wound D-string along $\tilde S^1$. This
D-string is in general broken up at D3's, and the open D-string
segments between adjacent D3's can move freely along $R^3$
directions. These open D-string segments can be
interpreted as fundamental monopoles~\cite{piljin}.

As there are $N$ such intervals, there are $N$ fundamental monopoles,
corresponding to the roots in the extended Dynkin diagram. These $N$
solitons constitute an instanton on $R^3\times S^1$, also known as a
caloron~\cite{piljin,lu,kraan,nahm}. The corresponding Nahm data encodes
the positions of the D-string segments and gauge fields on them.

Suppose we introduce noncommutativity on $R^3 \times S^1$, by turning
on a uniform NS-NS tensor field on it; 
\begin{equation}
[x_\mu,x_\nu]= i\theta_{\mu\nu}, \qquad i=1,2,3,4.
\end{equation}
The antisymmetric 2-tensor $\theta_{\mu\nu}$ will be assumed to be
covariantly constant. The ADHM formalism of the instanton in the
noncommutative Yang-Mills theory is developed in
Ref.~\cite{nek,nakajima}, where it was shown that the ADHM equation
(or D-term condition) acquires a triplet of constant terms that can be
thought of as Fayet-Iliopoulos term of the D0 worldvolume theory in
the presence of D4-branes. This triplet of number $\zeta_a$'s
($a=1,2,3$) are related to the anti-self-dual part of $\theta$ as
\begin{equation}
\theta^{(-)}=\zeta^a
\left(dx^4\wedge dx^a-\frac{1}{2}\epsilon^{abc}dx^b\wedge dx^c\right).
\end{equation}

Because we compactified one direction, the ADHM system actually consists of
infinite number of mirror images of D0. T-dualizing this picture to D3-D1
system, the D-term conditions become one-dimensional anti-self-dual equations
known as Nahm's equation~\cite{nahm}:
\begin{equation}
\frac{dT^a}{dt}+\epsilon_{abc}[T^b,T^c]=\zeta^a+\sum_{i=1}^{N}\delta(t-t_i)
\;a_i\sigma^a a_i^\dagger .
\end{equation}
The coordinate $t$ parameterizes the circumference of the dual circle
$\tilde S^1$, and $t=t_i$'s are positions of D3-branes along $\tilde
S^1$. We have chosen the coordinate $t$ to have dimension of mass,
while $T^a$'s have dimension of lengths, by inserting appropriate
factors of $\alpha'$, so that $0\le t \le 2\pi \tilde{R}/\alpha'$.
$T^a$'s are $k\times k$ unitary matrices where $k$ is the Pontryagin
number. The first term on the right hand side comes from the
noncommutativity, while the second terms reflects the possibility that
D-strings are broken up along each D3. Each $a_i$ is a
$2k$-dimensional complex vector $a_i^{\:A\alpha}$, $A=1,\dots,k$,
$\alpha=1,2$, and $\sigma^a$'s are the usual Pauli
matrices. Conventional Nahm's equation for calorons is recovered when
$\zeta^a$'s are taken to zero.

For a single instanton with $k=1$ (a self-dual instanton), the numbers
$T^a$ encode positions of D-string segments along $R^3$ directions and
so the interpretation of $\zeta^a$'s are quite clear. In the
coordinates $t$ and $x^a$'s, all D-string segments are slanted with
the slope of $\vec\zeta$, as in figure 1. This simple picture contains
much of physics we need to understand about instantons and monopoles
in noncommutative Yang-Mills theories. (For another view of the
slanted strings between D-branes, see Ref.~\cite{hash}.)

The D-string segments are magnetically charged with respect to the
(unbroken) $U(1)$ worldvolume gauge fields on $D3$'s, and are nothing
but non-Abelian magnetic monopoles. Clearly the single instanton on
$\tilde S^1\times R^3$ consists of a collection of D-string segments
that completes a singly-wound D-string, and hence a collection of $N$
distinct monopoles. Interestingly enough, the size $\rho$ of the
instanton is related to another measure of distance in the
multi-monopole picture as
\begin{equation}
\rho^2  =2R  \,\left(|\vec x_1-\vec x_{2}|+|\vec x_2-\vec x_{3}|+\cdots
+|\vec x_{N-1}-\vec x_N|+|\vec x_N-\vec x_{1}|\right).
\end{equation}
The parameter $\rho$ defined this way coincides with the conventional
definition of instanton size in $R^4$  when $\rho$ is much smaller
than $ R$. 

\vskip 2cm

\begin{center}
\leavevmode
\epsfysize=3in
\epsfbox{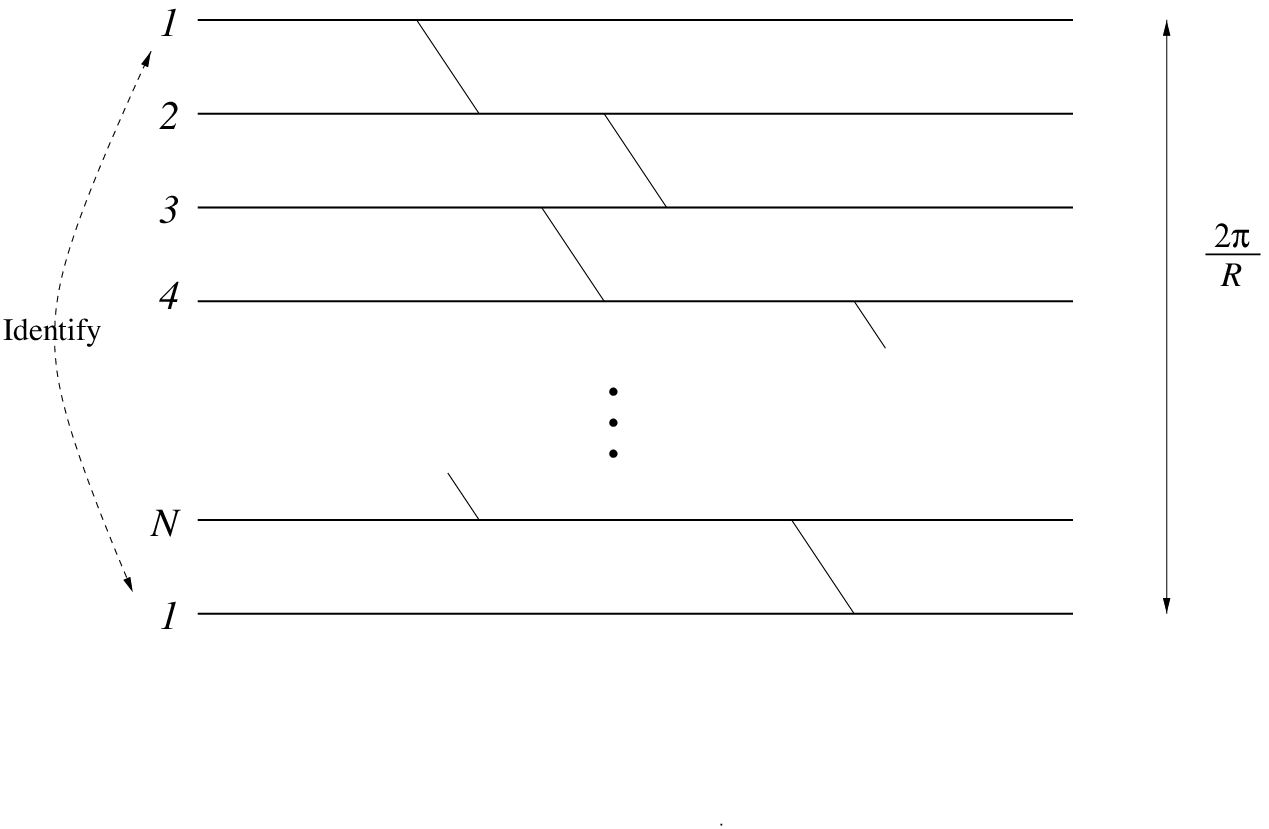}
\end{center}
\begin{quote}
{\bf Figure 1:} $N$ D3-branes, parallel to $R^{1+3}$, are represented by
horizontal lines. They are located at specific points along $\tilde S^1$,
whose radius is $1/R$. The D-string segments ending on D3-branes
are slanted toward the direction given by $\vec \zeta$.
\end{quote}

One immediate difference between the commutative and the noncommutative cases
is that the right hand side can vanish only for the former.  In
particular, the instanton on a single D4-brane is a $U(1)$ instanton,
yet its size is not zero; the noncommutativity blows up the small
instanton to be of finite size. For general $N$, the smallest possible
value the right hand side can take is obtained when D-string is
connected at $N-1$ of $N$ D3-branes. The right hand side is then,
\begin{equation}
2R\times |2\pi \tilde R\vec \zeta/\alpha'|,
\end{equation}
where $\tilde R$ is the radius of $\tilde S^1$. Using the T-dual relationship
$\tilde R R/\alpha'=1$, we find that
\begin{equation}
\rho^2_{\rm minimum}= 4\pi | \vec \zeta|,
\end{equation}
which persists as we decompactify $ S^1$ to go back to $R^{1+4}$.

Note that the two endpoints of a D-string segment is located at
different points along $R^3$. Each endpoint is perceived as a magnetic
charge with respect to an unbroken $U(1)$ associated with the D3-brane
on which the endpoint is located, so this means among other things
that the classical solution associated with the segment is not
traceless, and so cannot be thought of as an $SU(N)$ configuration. It
is necessarily a $U(N)$ configuration. When $N=1$, the D-string has
one connected component but its two endpoints are located at two
different points on the D3, in particular. A $U(1)$ instanton on a
noncommutative $R^3\times S^1$ is a magnetic dipole~\cite{nek}. (
See Ref.~\cite{mono} for the explicit construction of the dipole term
for $U(2)$ BPS monopoles.)

\section{Moduli Space of a Single Instanton on $R^3 \times S^1$}

The moduli space of a unit $U(N)$ periodic instanton has been derived
previously in the context of ordinary Yang-Mills theory on $R^3 \times
S^1$. With a generic Wilson line, the instanton actually consists of $N$
distinct monopoles, the sum of whose magnetic charges vanish.
The moduli space of distinct monopoles are well-understood, and one finds
the following hyperKaehler metric for the $4N$ dimensional moduli
space~\cite{piljin}.
\begin{equation}
g_{\rm total}
=\frac{4\pi^2 R}{e^2}\,\left(
M_{ij}d\vec x_i\cdot d\vec x_j + (M^{-1})_{ij} (d\xi_i+\vec v_{ik}
\cdot d\vec x_k)(d\xi_j+\vec v_{jm} \cdot d\vec x_m)\right).
\end{equation}
$\vec x_i$ are $N$ three-vectors, while $\xi_i$ are periodic in $2\pi$.
The symmetric matrix $M_{ij}$ depends on differences, $\vec r_{ij}=
\vec x_i-\vec x_j$, only, and has the form,
\begin{equation}
\left( \begin{array}{cccccc}
\mu_1+1/r_{N1}+1/r_{12} & - 1/r_{12}& 0&  \cdots & 0 & - 1/r_{N1} \\
- 1/r_{12}& \mu_2+1/r_{12}+1/r_{23} & -1/r_{23} &  \cdots & 0 & 0\\
0 &  -1/r_{23} &  \mu_3+1/r_{23}+1/r_{34} &\cdots &0&0 \\
\cdots &\cdots &\cdots &\cdots &\cdots &\cdots \\
\cdots &\cdots &\cdots &\cdots &\cdots &\cdots \\
\end{array}\right)
\end{equation}
The quantities $\mu_i=2\epsilon_i/R \ge 0 $, with $\sum_i \epsilon_i=1$,
parameterize the Wilson line, or alternatively the relative positions of
D3-branes along the dual circle $\tilde S^1$. The vector potentials
$\vec v_{ij}$ are related to the scalar potentials in $M_{ij}$ by,
\begin{equation}
\vec \nabla_i M_{jk}=\vec \nabla_i \times \vec v_{jk}, \label{curl}
\end{equation}
which is necessary for the metric to be hyperKaehler.

One way to derive this metric is to consider dynamics of
well-separated $N$ monopoles, as in Ref.~\cite{piljin,moduli}. The
above metric captures the long-range electromagnetic and scalar
interactions precisely. In cases of distinct monopoles, no
short-distance correction arises, and this simple-minded derivation
gives the right answer.  Alternatively, one could start with the Nahm
data of a caloron, and derive the corresponding metric in the space of
the Nahm data. The above form of the metric was first conjectured by
the authors, and subsequently proven by Kraan using Nahm
data~\cite{kraan1}, where the moduli space is obtained by a finite
hyperKaehler reduction of some flat quaternionic space.

The latter method can be easily generalized to noncommutative cases,
since the only modification is via the FI constants $\zeta^a$'s. For
a single instanton, all that happens is that the moment maps are
shifted by an amount proportional to $\zeta^a$'s. Despite the complicated
nature of noncommutative Yang-Mills theories, thus, the modification of
the instanton moduli space is rather simple.
With this in mind, let us consider how noncommutativity may alter the
moduli space metric. 

Since the long distance behavior of noncommutative 
theories should be identical to that of the commutative version,
we expect the former method to be also informative for well-separated
monopoles. So let us concentrate on long-distance interaction
between D-string segments.
The $i^{th}$ D-string segments appears on $R^3$ as a pair of particles,
separated by a fixed vector, $(t_{i+1}-t_i)\vec\zeta$.
They are charged, respectively, positively with respect to the $i^{th}$ and
negatively with respect to the $(i+1)^{th}$ unbroken $U(1)$ on D3-branes; 
the net electric charge is $(0,,\dots,0,1,-1,0,\dots,0)$.  From this, 
we can see easily that the long
range interaction is modified only through the correction to  $M_{ij}$
(and thus to $\vec v_{ij}$) that shifts the harmonic functions
$1/|\vec x_i-\vec x_{i+1}|$ to,
\begin{equation}
\frac{1}{ |\vec x_i-\vec x_{i+1}| }\quad\rightarrow\quad
\frac{1}{|(\vec x_i+(t_{i+1}-t_i)\vec\zeta/2)-
(\vec x_{i+1}-(t_{i+2}-t_{i+1})\vec\zeta/2)|} .
\end{equation}
This harmonic function encodes all long-range interactions via massless
fields on the $(i+1)^{th}$ D3-branes.

On the other hand, the first $N-1$ such shifts can be absorbed in the 
definition of the Cartesian coordinate $\vec x_i- \vec x_{i+1}$'s. By doing
so, the only vestige of this deformation survives in $|\vec x_N-\vec x_1|$.
The modified metric is then identical to the above except that a scalar
potential in $M$ is modified in the following manner;
\begin{equation}
\frac{1}{|\vec r_{N1}|} \quad\rightarrow \quad \frac{1}{|\vec r_{N1}-
2\pi\vec\zeta/R|}=\frac{1}{|\vec x_N-\vec x_{1}-
2\pi\tilde R\vec\zeta/\alpha'|},
\end{equation}
while the rest remains unchanged
\begin{equation}
\frac{1}{|\vec r_{i,i+1}|}\quad\rightarrow \quad \frac{1}{|\vec r_{i,i+1}|}
=\frac{1}{|\vec x_{i}-\vec x_{i+1}|},\qquad i=1,\dots,N-1 .
\end{equation}
The vector potentials in $\vec v_{ij}$ are modified according to
the relationship in Eq.~(\ref{curl}).

One may separate out the noninteracting center-of-mass degrees of freedom
by introducing new coordinates,
\begin{eqnarray}
\vec x_{cm} &\equiv& \frac{\sum \mu_i\vec x_i}{\sum \mu_i},\\
\vec r_1 &\equiv& \vec r_{12},\\
&\vdots&\\
\vec r_{N-1}&\equiv& \vec r_{N-1,N},
\end{eqnarray}
and similar redefinitions for canonical conjugate momenta of $\xi_i$'s,
\begin{eqnarray}
\frac{\partial}{\partial\xi_{cm}}&\equiv &
\left(\sum \mu_i \frac{\partial}{\partial \xi_i}\right)/\sum \mu_i ,\\
\frac{\partial}{\partial \psi_1}&\equiv &
\frac{\partial}{\partial \xi_1}- \frac{\partial}{\partial \xi_2},\\
&\vdots& \\
\frac{\partial}{\partial \psi_{N-1}}&\equiv&
\frac{\partial}{\partial \xi_{N-1}}-\frac{\partial}{\partial \xi_{N}}.
\end{eqnarray}
In the new coordinate system, $\vec x_{cm}$ and $\xi_{cm}$ decouple from
the rest. The $4(N-1)$ dimensional interacting part of the moduli space
is given by the metric,
\begin{equation}
g=\frac{4\pi^2 R}{e^2}\,\left(
C_{AB}d\vec r_A\cdot d\vec r_B + (C^{-1})_{AB} (d\psi_A+\vec \omega_{AC}
\cdot d\vec r_C)(d\psi_B+\vec \omega_{BD} \cdot d\vec r_D)\right) .
\end{equation}
$\psi_A$ are periodic in $4\pi$, and the symmetric matrix $C_{AB}$ has the
form,
\begin{equation}
C_{AB}=\left(\mu_{AB} + \frac{\delta_{AB}}{|\vec r_A|} + \frac{1}{
|\sum_{A=1}^{N-1} \vec r_A - 2\pi\vec\zeta/R\;|}\right) .
\end{equation}
The last term is common to all
components. The vector potentials $\vec \omega_{AB}$ are again related
to $C_{AB}$ by
\begin{equation}
\vec \nabla_D \,C_{AB}=\vec \nabla_D \times \vec \omega_{AB} .
\end{equation}
The ``reduced mass matrix'' $\mu_{AB}$ is defined by the formula,
\begin{equation}
\sum_i \mu_i d\vec x_i^2 = (\sum_i \mu_i)\,d\vec x_{cm}^2 +\sum_{A=1}^{N-1}
\sum_{B=1}^{N-1} \mu_{AB}
d\vec r_A\cdot d\vec r_B .
\end{equation}
If and only if $\vec \zeta/R$ is nonzero, the relative moduli space is
smooth. Otherwise, there exists a singularity at origin,
$\vec{r}_A=0, A=1,\dots,N-1$. 

While we derived this metric from the asymptotic interactions between
the D-string segments, we have many reasons to believe that we actually
found the exact metric. First of all, known moduli spaces of distinct
monopoles are always such that long distance dynamics determine the
metric everywhere.  Furthermore, the shift of moment maps one would
have considered in Nahm data approaches is exactly the shift of 
coordinate $\vec x_i$'s. In the following section, we will proceed to
solve for low energy dynamics of a single instanton with this moduli space.

\section{Low Energy Dynamics with or without Higgs Expectation}

The instanton soliton breaks half of the supersymmetry present in Yang-Mills
theory, and its low energy dynamics is given by the sigma model with
four complex supercharges~\cite{gaume};
\beq
{\cal L}={1\over 2} \left( g_{\mu\nu} \dot{z}^\mu \dot{ z}^\nu + i
g_{\mu\nu} \bar\eta^\mu \gamma^0 D_t \eta^\nu + {1\over 6}
R_{\mu\nu\rho\sigma}\bar\eta^\mu \eta^\rho \bar\eta^\nu \eta^\sigma\right)
\eeq
where we introduced a two-component real  fermionic coordinates
$\eta^\mu$ for each $\mu$.
Actually, this dynamics does not take into account possible Higgs expectation
values. We are already working in broken Coulomb vacua, due to the Wilson
line along $S^1$, so have no reason to exclude  adjoint
Higgs expectations.

Maximally supersymmetric Yang-Mills theory in five dimension is
written in terms of the vector multiplet, which consists of a vector
field, five scalar fields and a pair of Dirac fields in five dimensions.
Pictorially the five scalar fields encodes the fluctuation of D4- (or D3-)
branes along the five Euclidean directions transverse to $ R^{1+3}\times S^1$
(or to $R^{1+3}\times \tilde S^1$). 
In the presence of small expectation value of a single Higgs field, 
the low energy dynamics is modified by a potential term. For instance, if
one of the scalars gets a vev, the Lagrangian is corrected to~\cite{dongsu},
\beq
{\cal L}={1\over 2} \biggl( g_{\mu\nu} \dot{z}^\mu \dot{ z}^\nu +
g_{\mu\nu} \bar\eta^\mu \gamma^0  D_t \eta^\nu + {1\over 6}
R_{\mu\nu\rho\sigma}\bar\eta^\mu \eta^\rho \bar\eta^\nu \eta^\sigma
 - g^{\mu\nu} G_\mu G_\nu - D_\mu G_\nu  \bar\eta^\mu
\gamma_5 \eta^\nu \biggr),
\label{action}
\eeq
where the Killing vector field $G$ is defined through,
\begin{equation}
G=\sum a_i\frac{\partial}{\partial\xi_i},
\end{equation}
with eigenvalue $a_i$'s of the adjoint Higgs expectation in a suitable
normalization. Canonical quantization conditions 
enable one to translate the supercharges to geometrical operators on
the moduli space,
\begin{equation}
 Q \quad\rightarrow\quad d-i_G
\end{equation}
and similarly for its conjugate. The wavefunction is now represented
by differential forms on the moduli space. Here, $i_G$ denote the contraction
of the wavefunction/differential form with $G$. 
The SUSY algebra has the central charge,
\begin{equation}
Z\equiv i{\cal L}_G ,
\end{equation}
which measures the electric part of the energy,
and the Hamiltonian,
\begin{equation}
H= \frac{1}{2}\left(QQ^\dagger+Q^\dagger Q\right),
\end{equation}
is bounded below by the absolute value of the central charge $|Z|$.
In particular, the BPS bound states without any electric charge should
be annihilated by all supercharges.

\section{Counting Bound States with No Electric Charge}

Let us start with a single instanton of the $U(2)$ theory. This is the simplest
case with the nontrivial moduli space. The relative moduli space is
a 4-dimensional hyperKaehler space with double Taub-NUT centers.
\begin{equation}
g = U(\vec r)\,d\vec r^2 + U(\vec r)^{-1}(d\psi+\vec \omega\cdot d\vec r)^2,
\end{equation}
where
\begin{equation}
U(\vec r)=\left( \mu+\frac{1}{|\vec r|} +
\frac{1}{|\vec r -2\pi\vec \zeta/R\,|} \right ),
\end{equation}
and
\begin{equation}
\vec \nabla\times \vec\omega =\vec \nabla U.
\end{equation}
First, suppose that there is no scalar Higgs expectations, so that
$G\equiv 0$.  Then the problem of finding ground states reduces to
that of finding normalizable harmonic forms on the double-centered
Taub-NUT. The solution to such a problem is actually well-known for
arbitrary number of the Taub-NUT centers; For each Taub-NUT centers,
there exists precisely one associated (anti-self-dual) harmonic
2-form~\cite{ruback}. The harmonic forms can be written generally as,
\begin{equation}
d(f/U)\wedge (d\psi+\vec \omega\cdot d\vec r)-\frac{1}{2} 
U\epsilon_{ijk}\partial_i (f/U)
dr_j \wedge dr_k ,
\end{equation}
where $f=f(\vec r)$ with certain harmonic functions in $R^3$ spanned by
$\vec r$. Two regular and normalizable harmonic forms on the moduli space
are obtained by setting
by setting
\begin{eqnarray}
f= \frac{1}{|\vec r|}+\frac{1}{|\vec r -2\pi\vec \zeta/R\,|}
&\rightarrow &\Omega_1\\
f= \frac{1}{|\vec r|}-\frac{1}{|\vec r -2\pi\vec \zeta/R\,|}
&\rightarrow &\Omega_2
\end{eqnarray}
are regular and normalizable.
Call them $\Omega_1$ and $\Omega_2$, respectively 
 Note that the third obvious solution with $f=1$
is actually a constant multiple of $\Omega_1$.
Thus, when the Higgs expectation (beyond the Wilson line along $S^1$) is
absent, we find exactly two bound states with the unit Pontryagin number.

Both $\Omega_1$ and $\Omega_2$ consist of two lumps localized at
$\vec r=0$ and at $\vec r = 2\pi\vec \zeta/R$.
There is an illuminating
classical picture of these two lumps. The position $\vec r=0$ translates
to the statement that the two D-string segments are glued along  one of 
the D3-branes. Similarly $\vec r = 2\pi\vec \zeta/R$ represents D-string
segments glued along the other D3. Thus, the bound states at zero energy
are represented as linear combinations of two such glued D-strings.

In more general vacua with one or more Higgs expectation values, these glued
D-strings are classical bound states at $\vec r = 0 $ or $\vec r = 2\pi\vec
\zeta/R$ with net binding energy. For this reason, we expect that these
two states get deformed but remain as quantum bound states,
when the potential term due to adjoint Higgs expectations is turned on.

More generally, this suggests that there are $N$ independent bound
states with unit Pontryagin number in the  noncommutative
$U(N)$ theory on $S^1\times R^3$. Quantum mechanically, one should find
$N$ normalizable differential forms, satisfying a SUSY condition on
the $4(N-1)$ dimensional moduli space. This  needs a further work.  Here
we will be content  with finding the classical ground states of the
potential when we turn on one additional Higgs field.

Natural candidates for classical ground states are  D-string segments
glued at all except one D3-brane. 
Such classical states are represented by points on the moduli space,
where $N-1$ of the following $N$ vectors vanish,
\begin{equation}
\vec r_1, \vec r_2,\dots, \vec r_{N-1},
\vec r_N + 2\pi\vec\zeta/R .
\end{equation}
We introduced a new notation
\begin{equation}
\vec r_N \equiv - \sum_{A=1}^{N-1} \vec r_A .
\end{equation}
Does the bosonic potential vanish at such points?
For generic vev of a single Higgs, the nontrivial part of the bosonic
potential has the form~\cite{dongsu},
\begin{equation}
V=\frac{1}{2}\,a^A a^B (C^{-1})_{AB}
\end{equation}
This potential will vanish where all eigenvalues of $C$ diverge. Clearly,
the point where
\begin{equation}
\vec r_1=\vec r_2=\cdots=\vec r_{N-1}=0
\end{equation}
is one such ground state; All diagonal elements of $C$ diverges while all 
off-diagonal
elements are finite. On the other hand, one can choose a slightly different
coordinates $\vec r_A^{\, '}$ such that
\begin{eqnarray}
\vec r_1^{\, '}&=&\vec r_N+2\pi\zeta/R,\nonumber \\
\vec r_2^{\, '}&=&\vec r_1,\nonumber \\
\vdots \nonumber \\
\vec r_{N-1}^{\, '}&=&\vec r_{N-2},
\end{eqnarray}
which implies
\begin{eqnarray}
\vec r_{N}^{\, '}+2\pi\vec \zeta/R \equiv - \sum_A\vec r_A^{\, '}+
2\pi\vec \zeta/R &=& \vec r_{N-1}
\end{eqnarray}
When accompanied by a related transformation for the angular part,
this redefinition leaves the form of the metric invariant. In the new
coordinates,
\begin{equation}
\vec r_1^{\, '}=\vec r_2^{\, '}=\cdots=\vec r_{N-1}^{\, '}=0 ,
\end{equation}
is clearly a zero of the potential $V$. Repeating the exercise, we can see
that zeros of $V$ occur where any $N-1$ of $\vec r_1, \dots, \vec r_{N-1},
\vec r_N+2\pi\vec\zeta/R$ vanish. There are exactly $N$ such points. We
surmise that, in the noncommutative $U(N)$ theory on $R^{1+3}\times S^1$, 
there are exactly $N$ BPS supermultiplets of states with the
unit Pontryagin number.

In the limit $\vec \zeta=0$, the moduli space becomes singular. In the
Coulomb phase, at least part of $N$ states seems to survive. This can
be seen explicitly for the $U(2)$ case. In this limit, the two
Taub-NUT centers coalesce into one, and the relative moduli space
becomes an $Z_2$ orbifold of the single-center Taub-NUT. In the
process, $\Omega_2$ vanishes by itself (or disappear into the orbifold
point), while $\Omega_1$ remains finite and well-defined. This state
$\Omega_1$ is similar to the threshold bound states in monopole
dynamics~\cite{gaunt}.  It is not clear how many actually survive the
limit for general $N$, but it seems not farfetched to expect at least
one of them does. The turning on Higgs expectation instead of or in
conjunction with the Wilson line should not decrease the number of
states, and we expect to have at least one pure instanton state in the
Coulomb phase of the $U(N)$ theory.

\section{Decompactification}

One can take the decompactification limit by sending $R\rightarrow\infty$.
To reach a sensible moduli space metric in such a limit, we need to
rescale the relative coordinates $\vec r_A$ by
\begin{equation}
\vec y_A = R\, \vec r_A, \quad A=1,\dots ,N-1,
\end{equation}
and thus
\begin{equation}
\vec y_N\equiv - \sum_{A=1}^{N-1} \vec y_A =R\, \vec r_N,
\end{equation}
also. Upon such a rescaling the constant piece $\mu_{AB}$ gets multiplied by
$1/R$ and can be ignored. The remaining pieces are written as,
\begin{equation}
\frac{8\pi^2}{e^2}\left(\sum_A \left(
\frac{1}{y_A}d\vec y_A^2 + y_A\,(D\psi_A)^2\right)
+\frac{\left(\,\sum_A d\vec y_A\,\right)^2}{|\vec y_N+2\pi \vec \zeta|}
-\frac{\left(\,\sum_A y_A\,D\psi_A\,\right)^2}{|\vec y_N+2\pi \vec \zeta|
+ \sum_A |\vec y_A|}\right),
\end{equation}
where
\begin{equation}
D\psi_A\equiv d\psi_A+\vec \omega_{AB}(\vec r_C)\cdot d\vec r_B
= d\psi_A+\vec \omega_{AB}(\vec y_C)\cdot d\vec y_B,
\end{equation}
remains unchanged under the rescaling. All sums are for
$A=1,\dots,N-1$. The resulting metric is called the Calabi
metric~\cite{calabi}.

While the details of the dynamics have changed upon
$R\rightarrow\infty$, the ground state structure of the bosonic
potential, which becomes confining, did not. The bosonic potential
still vanishes at $N$ different points, where $N-1$ of $N$ vectors,
$\vec y_1$, $\vec y_2$, $\dots$, $\vec y_{N-1}$, $\vec y_{N}+2\pi\vec
\zeta$, vanish. Thus, as long as the theory is in the Coulomb phase,
there must be $N$ independent supermultiplets of states with a unit
Pontryagin number.

If one approaches the symmetric phase where the $U(N)$ gauge symmetry
is restored, the low energy effective potential disappears, and some
of the $N$ quantum states might disappear. In the case of $U(2)$, it
can be seen explicitly that $\Omega_1$ becomes nonnormalizable while
$\Omega_2$ remains normalizable.  It is unclear what physics is
responsible for such disappearance of some instanton states, or
whether this low energy phenomenon is meaningful at all in the full
Yang-Mills theory context.

For the case electrically charged case with the potential, there is a
recent work in the decompactified limit~\cite{lambert}. 

\section{Summary}

We explored the low energy dynamics of five-dimensional $U(N)$ Yang-Mills
theory in the noncommutative setting. One may consider the noncommutativity as
a convenient short-distance regulator that allows us to discuss the
quantum states of instanton solitons. We computed the moduli space metric
of a single $U(N)$ instanton, which is smoothed out thanks to the 
noncommutativity, and wrote down the supersymmetric low energy  dynamics
explicitly.

In the Coulomb phase, where the symmetry is broken to $U(1)^N$ by a
Wilson line or adjoint Higgs expectation values, there are exactly $N$
supermultiplets of states with the unit Pontryagin numbers and no
electric charge. States with electric charges in addition to the
Pontryagin number can be studied with the given low energy dynamics,
which we have  not attempt here.

\section*{Acknowledgment}

This work was initiated while one of the authors (P.Y.) was visiting
the Aspen Center for Physics. We are grateful to D. Bak, C.-S. Chu,
and K. Hori for useful conversations. K.L. is supported in part by the
SRC program of the SNU-CTP and by KOSEF 1998 Interdisciplinary
Research Grant 98-07-02-07-01-5.


\begin{thebibliography}{99}

\bibitem{connes}
A. Connes, M.R. Douglas, and A. Schwarz,
{\it Noncommutative geometry and Matrix theory: compactification on
tori}, hep-th/9711162, JHEP 9802 (1998) 003.


\bibitem{doug} M.R. Douglas and C. Hull, {\it D-branes and the
noncommutative torus}, hep-th/9711165, JHEP 9802 (1998) 008;
Y.-K. E. Cheong and M. Krogh, {\it Noncommutative geometry from 0
branes in a background B field}, Nucl Phys. B528 (1998) 185; C.-S. Chu
and P.-M. Ho, {\it Noncommutative open string and D-brane}, Nucl
Phys. B550 (1999) 151; C.-S. Chu and P.-M. Ho, {\it Constrained
quantization of open string in background B field and noncommutative
D-brane}, hep-th/9906192; V. Schomerus, {\it D-branes and deformation
quantization}, JHEP 9906:030 (1999); F. Ardalan, H. Arfaei and
M.M. Sheikh-Jabbari, {\it Mixed branes and M(matrix) theory on
noncommutative torus}, hep-th/9803067; F. Ardalan, H. Arfaei and
M.M. Sheikh-Jabbari, {\it Noncommutative geometry from strings and
branes}, hep-th/9810072, JHEP 02 (1999) 016;F. Ardalan, H. Arfaei and
M.M. Sheikh-Jabbari, {\it Dirac Quantization of open strings and
noncommutativity in branes}, hep-th/9906161.


\bibitem{seiberg}  N.  Seiberg and  E.  Witten, {\it String theory
and noncommutative geometry},  JHEP 9909 (1999) 032.



\bibitem{witten} E. Witten, {\it Small instantons in string theory},
Nucl. Phys. B460 (1996) 541; M.R. Douglas, {\it Branes with branes},
hep-th/9512077. 

\bibitem{nek}
N. Nekrasov and A. Schwarz, {\it Instantons on noncommutative $R^4$, and
(2,0) superconformal six dimensional theory},  Commun. Math. Phys. 198
(1998) 689.

\bibitem{ber} N. Berkooz, {\it Nonlocal field theories and the
noncommutative torus}, Phys. Lett. B430 (1998) 237.

\bibitem{nakajima}
H. Nakajima, {\it Resolutions of moduli spaces of ideal instantons on
$R^4$}, in {\it Topology, geometry and field theory}, (World
Scientific, 1994); M. R. Douglas and G. Moore, {\it D-branes, Quivers, 
and ALE Instantons}, hep-th/9603167.

\bibitem{adhm} M.F. Atiyah, N.J. Hitchin, V.G. Drinfeld and
Yu. I. Manin, Phys. Lett. 65 A (1978) 185.

\bibitem{piljin}
K. Lee and P. Yi, {\it Monopoles and instantons on
partially compactified D-branes}, Phys.Rev. D56 (1997) 3711.  


\bibitem{lu} K. Lee, Phys. Lett. B426 (1998) 323; K. Lee and C. Lu,
Phys.Rev. D58 (1998) 025011.

\bibitem{kraan}
T.C. Kraan and P. van Baal, Phys. Lett. B428 (1998) 268;
Nucl. Phys. B533 (1998) 627; Phys.Lett. B435 (1998) 389. 


\bibitem{nahm} W. Nahm, {\it Self-dual monopoles and calorons}, in
Lect. Notes. in Physics, 201, eds, G. Denardo, etc. (1984) p. 189.


\bibitem{hash} A. Hashimoto and K. Hashimoto, {\it Monopoles and dyons
in non-commutative geometry}, hep-th/9909202, JHEP 9911 (1999) 005.


\bibitem{mono} D. Bak, {\it Deformed Nahm equation and a
noncommutative BPS monopole}, hep-th/9910135; K. Hashimoto, Hiroyuki
Hata, and S. Moriyama {\it Brane configuration from monopole
solution in non-commutative super Yang-Mills theory}, hep-th/9910196.

\bibitem{moduli}
K. Lee, E.J. Weinberg and P. Yi, {\it Moduli Space of many BPS monopoles for
arbitrary gauge groups},  Phys. Rev. {\bf D54}, 1633 (1996).



\bibitem{kraan1} T.C. Kraan, {\it Instantons, monopoles and toric 
hyperkaehler manifolds}, hep-th/9811179; M.K. Murray, {\it A note  on
the (1,1,...,1) monopole metric}, hep-th/6505054, J. Geom. Phys. 23
(1997) 31. 


\bibitem{gaume} L. Alvarez-Gaume and D. Freedman, Commun. Math. Phys.
91 (1983) 87.

\bibitem{dongsu} D. Bak, C. Lee, K. Lee, and P.Yi, 
{\it Low energy dynamics for 1/4 BPS dyons}, hep-th/9906119;
{\it Quantum 1/4 BPS dyons}, hep-th/9907090.



\bibitem{ruback} 
P. Ruback, Commun. Math. Phys. 107 (1986) 93; A. Sen,
hep-th/9705212, Adv. Theor. Math. Phys. 1 (1998) 115.

\bibitem{gaunt} J.P. Gauntlett and D.A. Lowe, Nucl. Phys. B472 (1996)
194;  K. Lee, E.J. Weinberg, and P. Yi, Phys. Lett. B376
(1996) 97; G.W. Gibbons, Phys. Lett. B382 (1996) 53.


\bibitem{calabi} G.W. Gibbons, P Rychenkova and R. Goto, 
Commun. Math. Phys. 186 (1997) 581; 
E. Calabi, {\it Isometric families of k\"ahler strutures}, in The Chern
Symposium 1979, ed. W.Y. Hsiang et.al., Springer-Verlag, New York
(1980). 

\bibitem{lambert}
N.D. Lambert and D. Tong,
{\it Dyonic instantons in five dimensional gauge theories}, 
hep-th/9907014, Phys. Lett. B462 (1999) 89.


\end{thebibliography}
\end{document}